\def\ut#1{\mathop{\vtop{\ialign{##\crcr
     $\hfil\displaystyle{#1}\hfil$\crcr\noalign
     {\kern1pt\nointerlineskip}\hbox{$\hfil\sim\hfil$}\crcr
     \noalign{\kern1pt}}}}}
\def\undersim{\ut}   
\magnification=\magstep1
\tolerance=1700
\hoffset 0.truecm
\voffset 0.0truecm
\hsize 15.5truecm
\vsize 22.5truecm
\def\unomez{\vskip 1.5truecm}
\def\uno{\vskip 1truecm}
\def\due{\vskip 2truecm}
\def\mezzo{\vskip .5truecm}
\def\skip{\vskip 0.2truecm\noindent}
\def\skipi{\vskip 0.2truecm}
\baselineskip=14pt
\def\newline{\par\noindent}
\centerline{ {\it Paper presented at the 1996 Star Formation Workshop,}}
\centerline{{\it  Wellesley College, July 1996}}
\due
\centerline {\bf THE EVOLUTION OF}
\centerline {\bf EMBEDDED SMALL CLUSTERS}
\unomez
\centerline {R. Capuzzo Dolcetta, F. Sori}
\uno
\centerline {{\it Inst. of Astronomy, University "La Sapienza", Roma, Italy}}
\uno
A young open cluster is a {\it $2$-phase system}: 
\skip$\bullet$ an ensemble of {\bf stars} move in
a {\bf gaseous} medium (the mother molecular cloud).
\skipi The {\it dynamics} and {\it thermodynamics} of the system, and so its
\skip --{\it evolution} and {\it final fate} (is it stable or unstable?)
\skip {\bf strongly depends on the mutual feedback} between gas and stars.
\skip We present an approach which consists in a (simplified) model
where stars (N--bodies) move within a gaseous spherical molecular
cloud. The two components influence each other through 
\skip\centerline{-- {\it gravity} and {\it mass loss.}}
\skip\skip  Among other results (role of IMF, SFE, stellar background,
etc., see {\bf Conclusions}), 
we find that a significant fraction of small
clusters can be destroyed even
\skip\centerline {-- {\bf before} SN explosion.}
\skip when a {\it significant} amount of massive stars are present.
\mezzo
\centerline {\bf THE MODEL}
\skipi
After the Lada, Margulis and Dearborn (1984, ApJ 285, 141 LMD) work 
not much has been done
 to study {\it quantitatively}
the early evolution and fate of stellar clusters embedded in their
mother cloud, following numerically the
N--body dynamics of stars moving in a (dispersing) gaseous cloud. 
 The LMD model was not fully self--consistent, for
the gas was assumed to expand with an assumed time law; by the way, this
work gave relevant information on the capability of a stellar
system to  {\bf remain bound after gas removal} in dependence on the
star formation  efficiency (SFE).
\skipi -- An answer to the crucial question: 
\skip$\bullet$ {\it what conditions on IMF
and on SFE allow a small cluster, emerging from a molecular cloud,
to remain bound?}
\skip\skipi -- {\bf necessarily} implies that the mutual feedback 
between gas and stars is {\bf taken into account}.
\skip\skip To get really {\it reliable} results one should couple 
an {\bf N--body 
code} to a {\bf fully hydro--code} to model the radiative
and mechanical interaction between the {\it stellar} and {\it gaseous} phases. 
This has been partially done (see Capuzzo--Dolcetta
and Di Lisio, {\it SPH in Astrophysics}, 1994, Mem.S.A.It., 65, 1107), and 
is the target of future work (Capuzzo--Dolcetta, Di Lisio, Navarrini, Palla,
in prepaparation).
\par Results good to order of magnitude, can however be obtained
with the present model, which treats the coupled dynamics and thermodynamics 
of stars and gas in a cluster with the following:
\skip\centerline{\it \bf approximations}
\skip $\bullet$  the gas cloud evolves in time keeping {\it spherical shape}
and a spatially {\it uniform} density.
\skip $\bullet$ the gravitational force exerted by stars on the  cloud is 
approximated.
\uno
\skip\centerline{\bf The Equations}
\mezzo
The relevant equations are:
\skip  
$$
\cases{
 {\vec a}_i= {\displaystyle {\vec F} \over \displaystyle m_i} - 
 {\displaystyle dm_i \over \displaystyle dt} 
{\displaystyle {\vec v_i} \over \displaystyle m_i}
 & $i=1,\ldots,N$ \cr\cr
 \ddot R = \ddot R_g + \ddot R_p + \ddot R_* +\ddot R_{ml} +  \ddot R_{vr} 
&{}\cr\cr
 \dot U = \dot U_{p} + \dot U_{vr} + \dot U_{ml} + \dot U_{SN} &{}\cr
}
$$
\skip
which is a $6N+3$ order system submitted to the appropriate initial conditions.
\skip
-- ${\vec a}_i$ is the $i$--$th$ star's acceleration,
\newline 
-- $R$ is the gas--sphere radius,
\newline
-- $U$ is the gas internal energy.
\vfill\eject
%
\centerline {In the gas {\it motion} equation:}
$$
 \ddot R_g=  -{GM\over R^2}  ~~(self-gravity)
$$ 
$$  
\ddot R_p =  { {3 \gamma (\gamma-1) U_g} \over {M_g R_g}} 
 ~~(pressure~field) 
$$
$$ 
\ddot R_* = {\displaystyle 1 \over \displaystyle M} \sum_{i=1}^N 
 f_i  ~~(stellar-gravity)
$$
$$
\ddot R_{ml} =  \sum_{i=1}^N \left ( 1 - 2  {\displaystyle r_i^3 \over 
\displaystyle R^3 } \right ) \ddot R_{{ml}_i} ~~(stellar~mass-loss)
$$
$$ 
\ddot R_{{ml}_i} =  {1 \over M}  {\displaystyle {\dot m_i}} v_{{ml}_i}
$$
$$
\ddot R_{vr} = - {k_{vr}\over \displaystyle 
{M+M_*}} \left({ {\displaystyle {M \dot R +\sum_{i=1}^N m_i \dot r_i }}
}\right) ~~(viol.~relax.)
$$
$\bullet$  $f_i$ is an approximation of the force exerted by the 
$i$--th star on the gas cloud.
\newline
$\bullet$  $v_{{ml}_i}$ is the $i$--th star wind speed (taken from the literature).
\mezzo
\centerline {In the gas {\it energy} equation:}
$$\eqalign{ 
\dot U_p= &{9 \over 5} 
 \gamma (\gamma-1) 
 {\displaystyle {\dot R U} \over \displaystyle  R} ~~(pressure~heating)\cr
\dot U_{vr} = &{3 \over 5} k_{vr} {\displaystyle {M 
 \over {M + M_*} }} \dot R \left({M\dot R+\sum_{i=1}^N m_i \dot r_i }\right)
~(viol.~relax.) \cr
\dot U_{ml} = &{3 \over 5} M \dot R_g  \sum_{i=1}^N \left | \left 
( 1 - 2 {\displaystyle {r_i^3} \over \displaystyle R^3 } \right ) 
\ddot R_{{ml}_i} \right | ~(star~mass-loss) \cr
\dot U_{SN} = &\delta (t- t_{SN})  e_{SN} ~~(SN~contribution)\cr}
$$
\vfill\eject
\centerline{\bf Parameters of the models}
\skip
Stars are initially {\it uniformly} distributed in space and velocity
in a sphere of radius $R_{*0}$ with velocities to satisfy the given 
virial ratio (here assumed =1).
\skip\skip The relevant initial parameters are:
\mezzo
\centerline {\it {\bf Stars:}}
$$N~=~~number~of~stars$$
$$R_{*0}=~~initial~cluster~radius = 1~pc$$
$$IMF\propto m^{-\alpha},~~0.2 \leq m/M_{\odot} \leq 20$$
$$local~ SFE \equiv \varepsilon $$ 
$$ chemical~ composition = (X,Y,Z) = (0.7,0.27,0.025) $$
$$virial~ratio \equiv \nu_0 = {2\times kinetic~energies \over 
potential~energy} = 1 $$
\mezzo
\centerline {\it {\bf Gas Clump:}}
$$R_0~ = ~~initial~radius= R_{*0}$$
$$\rho_0~ = ~~initial~gas~density=500~M_\odot/pc^3$$
$$\dot R_0~ = ~~initial~collapse~velocity~=0$$
$$U_0~ = ~~initial~internal~energy$$
\vfill\eject
\centerline {\bf CONCLUSIONS}
\skip$\bullet$ A small cluster ($N 
\undersim< 100,~0.2 \leq m/M_{\odot} \leq 20$) embedded in a gas clump 
of typical density 
$\rho \simeq 500M_{\odot}/pc^3$ 
\skip\centerline {is {\bf lost} to the background }
\skip at the time $t \simeq 10 Myr$ when first SN explode (see {\it Fig. 1}), 
if SFE$ \undersim< 0.4$ .
\skip$\bullet$ When SFE$ > 0.4$, the cluster {\bf resists} to the 
explosion whenever the IMF is {\bf not biased} towards {\bf large masses}.
\skip$\bullet$ When the exponent of the IMF$ \propto m^{-\alpha}$ is 
{\bf sufficiently negative} ($\alpha \undersim< -2$) and SFE$\undersim> 0.35$,
the gas cloud and its embedded cluster are {\bf disrupted by powerful stellar 
winds} in a shorter time (few $Myrs$). In this case,
$$\Downarrow$$
just {\bf very high} SFE allow the cluster to survive ({\it Fig. 2}, 
{\it Fig. 3} and {\it Fig. 4}).
\skip$\bullet$ The capability to distinguish a small cluster over a background
{\bf strongly depends} on the {\it cut-off} magnitude (the density contrast 
falls of a factor 50 when $V_{cut}$ is changed from 14 to 18 !). This means 
that {\it intrinsecally bound} cluster can be {\bf misconsidered as unbound} 
just because they are observed over a too crowdy background:
$$\Downarrow$$
so any practical definition of cluster lifetime {\bf must} take into account 
the bachground over which the cluster is projected ({\it Fig. 5} and 
{\it Fig. 6}).
\vfill\eject
\centerline{ {\bf Figure Captions} }
\unomez
\noindent
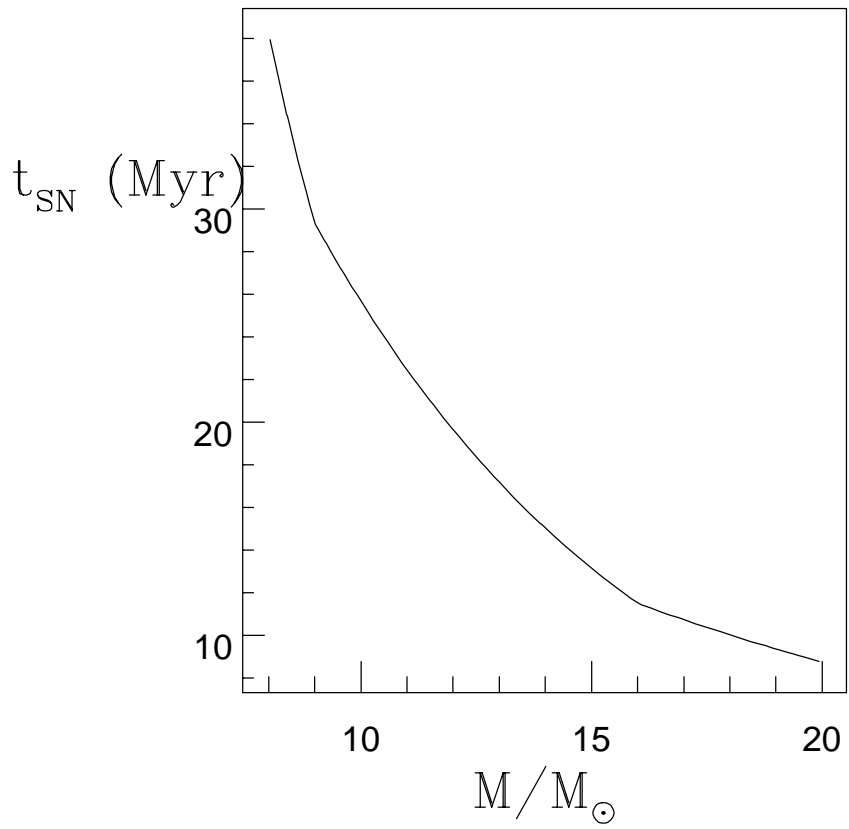
{\bf Fig. 1} : Super--Nova explosion time vs star mass (chemical composition 
$ X=0.7,~Y=0.27,~Z=0.025$).
\mezzo
\noindent
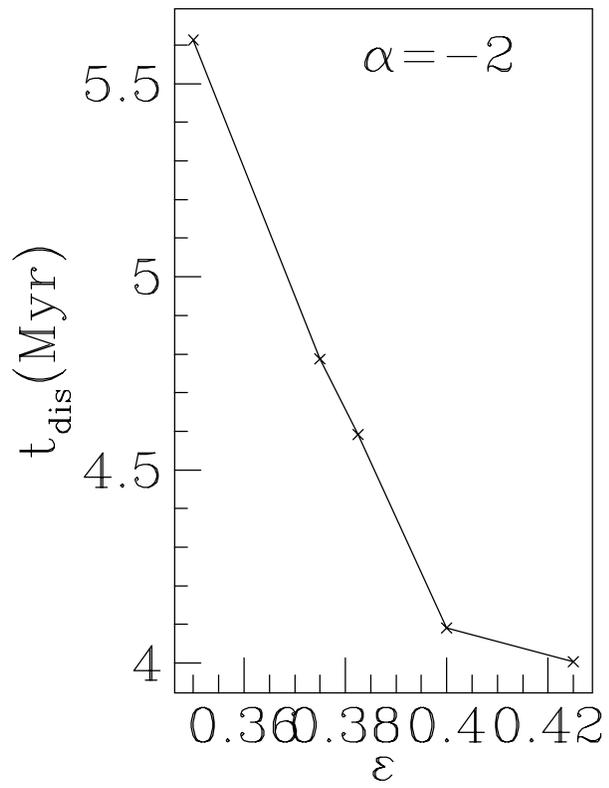
{\bf Fig. 2} : Gas cloud dissolution time vs SFE for the IMF exponent 
$\alpha=-2$.
\mezzo
\noindent
{\bf Fig. 3} : Rough delimitation of regions of bound and unbound 
clusters in the
($\alpha$,$\varepsilon$) plane, where $\alpha=$ IMF exponent, 
$\varepsilon =$SFE.
\par\noindent Values of $\varepsilon$ above the upper horizontal line
have not yet been investigated.
\mezzo
\noindent
{\bf Fig. 4} : For the models whose $\alpha,$ $\varepsilon,$ N are labelled 
on the top :
\par\noindent
{\it solid} lines --- cluster Lagrangian radii of 25\%, 50\%, 75\% and 100\% 
of the mass
\par\noindent
{\it dashed} lines - - gas cloud radii.
\par\noindent
Bottom panels are the enlargements of the upper ones.
\mezzo
\noindent
{\bf Fig. 5 a,b} : Time evolution of the cluster density contrast $\Delta \rho / 
\rho_{bg} = (\rho - \rho_{bg}) / \rho_{bg}$. $V_{cut}$ is the lower 
luminosity cut--off
of the background, which is taken at latitudes $b=0^\circ$ and $b=45^\circ$ 
(upper and lower curves, respectively, in each panel). Horizontal 
lines correspond to 
$\Delta \rho / \rho_{bg} =1$, below which the cluster is undistinguishable 
over the estimated background. The various cases studied label the panels.
\mezzo
\noindent
{\bf Fig. 6} : Enlarged view of part of Fig. 5b, to show
clearly the transition from "visible" to "unvisible" cluster.

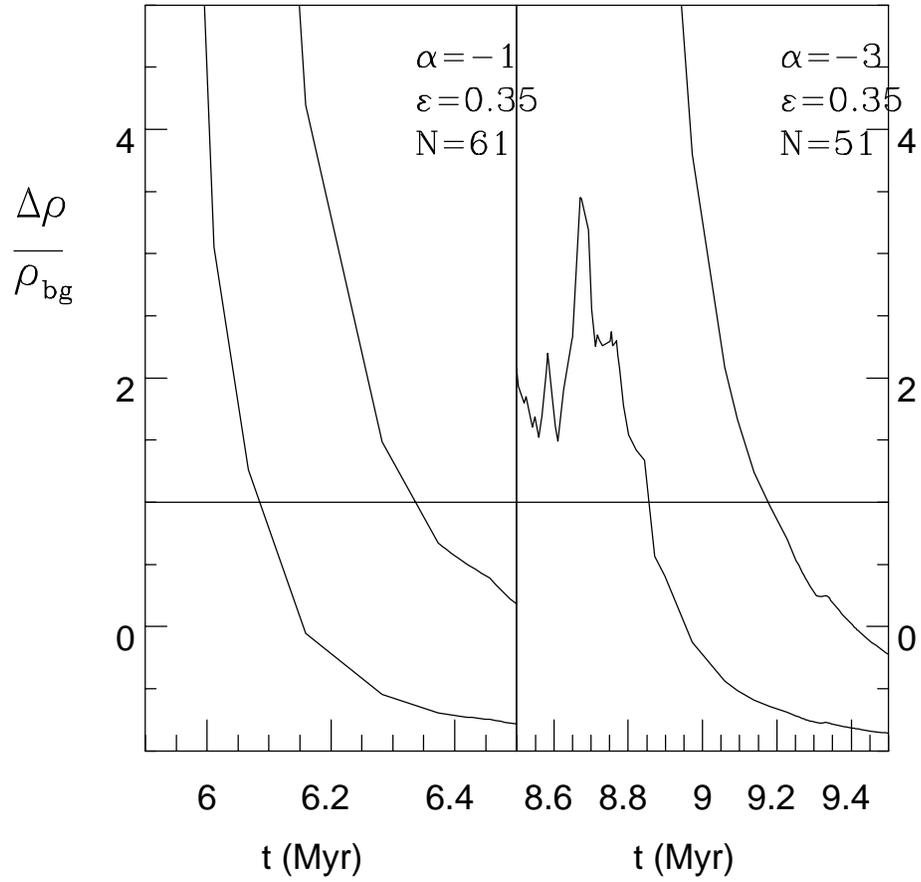
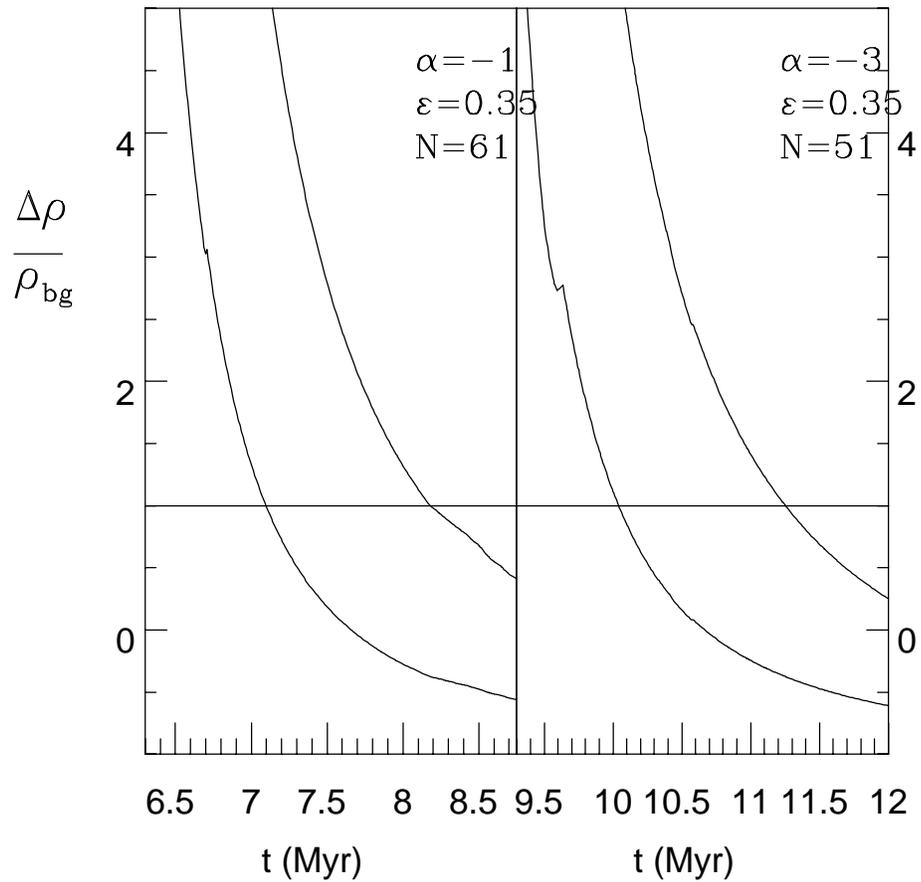
\bye